\documentclass[twocolumn,prd,superscriptaddress,preprintnumbers,amsmath,amssymb,amsmath,nofootinbib]{revtex4}

\usepackage{graphicx}
\usepackage{epsf}
\usepackage{epsfig}
\usepackage{cancel}

\usepackage{subfigure}
\usepackage{amsmath,amssymb}
\usepackage{rotating}
\usepackage{bm}
\usepackage{color}

\usepackage{verbatim}

\def\Sw{S}

\def\Sc{S_{\rm cl}}
\def\Lc{\mathcal{L}_{\rm cl}}
\def\Norm{{\cal N}}

\def\s0#1#2{\mbox{\small{$ \frac{#1}{#2} $}}}
\def\0#1#2{\frac{#1}{#2}}

\def\eq#1{(\ref{#1})}
\def\Eq#1{eq.~(\ref{#1})}

\def\ep{\epsilon}

\DeclareMathOperator{\Sdet}{\rm SDet}
\DeclareMathOperator{\Str}{\rm STr}

\def\M{M}
\newcommand{\Mc}{\tilde{M}}

\newcommand{\dec}{\tilde{\delta}}

\newcommand{\half}{\frac{1}{2}}
\newcommand{\bpsi}{\bar{\psi}}

\newcommand{\alm}{\alpha^{-1}}

\newcommand{\al}{\alpha}

\newcommand{\pa}{\partial}

\def\Tr{{\rm Tr}}

\newcommand{\R}{\mathcal{R}}
\newcommand{\Normw}{\Norm_{\mathcal{R}}}

\begin{document}

\title{Blocking-inspired supersymmetric actions: a status report}

\author{Georg Bergner}
\affiliation{Institut f\"ur Theoretische Physik,
Universit\"at Frankfurt,
D-60438 Frankfurt, Germany} 
\author{Falk Bruckmann}
\affiliation{Institut f\"ur Theoretische Physik, Universit\"at
  Regensburg, D-93040 Regensburg, Germany}
\author{Yoshio Echigo} 
\affiliation{Graduate School of Science and Technology,  
 Niigata University, 
{950-2181}, Niigata, Japan}
\author{Yuji Igarashi } 
\affiliation{Faculty of Education, 
 Niigata University, 
{950-2181}, Niigata, Japan}
\author{Jan M.~Pawlowski}
\affiliation{Institut f\"ur Theoretische
  Physik, Universit\"at Heidelberg, Philosophenweg 16, 69120
  Heidelberg, Germany}
\affiliation{ExtreMe Matter Institute EMMI, GSI
  Helmholtzzentrum f\"ur Schwerionenforschung mbH, Planckstr. 1,
  D-64291 Darmstadt, Germany}
\author{Sebastian Schierenberg}
\affiliation{Institut f\"ur Theoretische Physik, Universit\"at
  Regensburg, D-93040 Regensburg, Germany}

\begin{abstract}
  We provide a status report on the advances in blocking-inspired
  supersymmetric actions. This is done at the example of interacting
  supersymmetric quantum mechanics as well as the Wess-Zumino
  model. We investigate in particular the implications of a nontrivial
  realisation of translational symmetry on the lattice in this
  approach. We also discuss the locality of symmetry generators.

\end{abstract}

\maketitle

\section{Introduction}
\label{sec:intro}

There has been an increasing interest in lattice simulations of
supersymmetric theories as potential high energy  
extensions of the standard model. As any quantum field theory
supersymmetry (SUSY) has to be regularized, and the space-time lattice
is an obvious nonperturbative choice. On the lattice, however, SUSY is plagued by different
conceptual and practical problems, ranging from the explicit breaking
of supersymmetry by the boundary conditions to the violation of the
Leibniz and chain rule of differentiation by lattice derivative
operators \cite{Dondi:1976tx}. For the current status of lattice SUSY see, e.g.,
\cite{Catterall:2010jh} and references therein.

In Ref.~\cite{Bergner:2008ws} three of us suggested to break supersymmetry
in a controlled way similarly to the implementation of chiral symmetry for
Ginsparg-Wilson (GW) fermions \cite{Ginsparg:1981bj}. This approach
uses blocking as well-known from the renormalization group
context. Rather than trying to evaluate the corresponding effective
action, the concept of \cite{Bergner:2008ws,Ginsparg:1981bj} is to
focus on the modified symmetry obeyed by it. This leaves us with a
symmetry relation, or Ward-Takahashi identity, corresponding to a modified symmetry
transformation. It ensures the full symmetry in the continuum
limit, i.~e.\ in the limit of removing the regulator. Other
solutions than the one from blocking, which is typically unknown,
should be possible for it.

The aim of this approach is to represent SUSY on the lattice in a
similar way as a solution of such a symmetry relation
(\Eq{eq:SymWilson} below). 
The breaking of the Leibniz rule by any lattice difference
operator forbids the realization of the unmodified complete 
SUSY on the lattice \cite{Bergner:2009vg}. This is similar to the Nielsen-Ninomiya theorem 
in the case of chiral symmetry. Like in the Ginsparg-Wilson relation \cite{Ginsparg:1981bj} the nontrivial right 
hand side of the relation should account for this unavoidable breaking. 
One of the main obstacles in finding solutions is the obviously nonlinear character of
this relation for truly interacting, i.~e.\ higher than quadratic,
systems. The appearance of polynomial interactions of degree higher
than in the original theory or even non-polynomial interactions is
familiar from effective actions. Their contribution vanishes in the continuum limit.

Another aspect that needs to be investigated in this approach is locality of the resulting action.
Like the polynomial form, also locality might be lost in a generic effective action, e.~g.\ with a sharp momentum cut-off 
in the regulator. Nevertheless the locality in terms of the exponential suppression of the interaction with the distance is 
considered a basic requirement of a lattice theory. 
Note that in \cite{Bonini:2005qx} a supersymmetric lattice action with a similar modified locality 
was obtained using a different approach that implements nonlinear transformations on the lattice.

It may help to review conventional lattice actions from this point of
view. Non-abelian lattice gauge actions with local gauge invariance are
well-established. Only in some special cases these actions are derived from an 
explicit solution of the blocking from the continuum. A solution to the 
GW relation for chiral symmetry is the overlap operator, which can be obtained from
five-dimensional domain wall fermions. These solutions are derived according to the 
symmetry relation of the blocked action, but not as a solution of the blocking. Hence the 
symmetry relation is enough to ensure the symmetric continuum limit.
Note that for chiral symmetry the gauge fields are spectators, such that this action should be viewed as
quadratic in the fermion fields. Therefore, the derivation is simple compared to the generic case.

Solutions of the general symmetry relation have been worked out in
\cite{Bergner:2008ws} for free field theories and constant fields in
supersymmetric quantum mechanics. Here, we further pursue this approach. 
We discuss further properties of the modified symmetry relation and its solutions. We
give a status report on the advances made so far and an assessment of
the remaining obstacles.

 The present work is structured as follows. In Section~\ref{sec:setup}
we recall the formalism of \cite{Bergner:2008ws}. It is extended by an alternative way to
derive solutions in the trivial non-interacting case. The supersymmetric 
quantum mechanics is considered as an simple application of this general setup.  In
Section~\ref{sec:blocking_continuum} we discuss applications of the present
formalism to the continuum theory in the presence of a controlled
supersymmetry breaking. This includes in particular the
two-dimensional Wess-Zumino model. In Section~\ref{sec:translation} we
discuss necessity of the breaking of translational invariance in the
present setting. In Section~\ref{sec:sum} we summarise
our findings.

\section{The set-up}
\label{sec:setup}

\subsection{Blocking and the symmetry relation}
\label{sec:blocking}

In this section we briefly review the symmetry relation for arbitrary
linear symmetries that has been obtained via blocking in
\cite{Bergner:2008ws}. We consider some continuum theory with fields
$\varphi^i(x)$ and classical action $\Sc[\varphi]$, whose generating functional
reads,
\begin{eqnarray}\label{eq:genfunc1}
  Z[J]=\0{1}\Norm\int d\varphi \, e^{-\Sc[\varphi]+
    \int \!dx\,J^i(x) \varphi^i(x)} 
  \,, 
\end{eqnarray}
where the index $i$ sums over internal structures, Lorentz
indices and species of fields. In the following we consider theories
with a linear symmetry, to wit
\begin{eqnarray}\label{eq:classvar}
 \varphi\to\varphi+\dec\varphi\,,\quad 
 (\dec\varphi)^i(x) = \ep \int \! dy \,\Mc^{ij}(x,y)\varphi^j(y)\,,
\end{eqnarray}
where $\Mc$ may mix different field species, in the case of SUSY it does mix fermions and bosons.

A renormalization group step with the regulator $\R$ leads to the Wilsonian effective action $\Sw[\phi]$:
\begin{eqnarray} \label{eq:SWilson} 
e^{-\Sw[\phi]}:=\Normw\, \int d\varphi\,
  e^{-\R[\varphi,\phi]
  -\Sc[\varphi]}\,.
\end{eqnarray}
where in general $\phi^i$ are the fields of the regularized effective action.
We consider a specific quadratic regulator that is defined as
\begin{eqnarray} \label{eq:BlockRegulator}
\R[\varphi,\phi]&:=&\frac{1}{2}(\phi-\Phi[\varphi,f])_n^i \alpha^{ij}_{nm}(\phi-\Phi[\varphi,f])^j_m\\
\Normw&:=&\Sdet{}^{1/2}\alpha\, .
\end{eqnarray}
In our approach the blocked fields $\phi^i_n$ are the fields on the lattice. 
Therefore, the regulator involves as a first step an averaging of the fields with an
averaging function $f$ around a
lattice point $an$,
\begin{eqnarray}\label{eq:average} 
  \Phi[\varphi,f]^{i}_n:=\int d^d x\, f^{ij}(an-x)\, \varphi^j(x)\,, 
\end{eqnarray}
where $a$ is the lattice spacing and $n$ are integers labelling the
lattice sites.  $f$ may mix the different fields.
The blocking kernel $\alpha$ 
is assumed not to mix bosons and fermions. SDet is the
super-determinant, i.e. the determinant for bosons and its inverse for
fermions. When $\alpha\to\infty$, the lattice fields $\phi$ are
forced to be equal to the averaged fields $\Phi$ (absorbing the
superdeterminant) and \Eq{eq:SWilson} becomes
\begin{eqnarray} \label{eq:SWilsondelta} 
e^{-\Sw[\phi]}=\int d\varphi\,
  \prod_{i,n} \delta(\phi^i_n-\Phi[\varphi,f]^i_n)
  e^{-\Sc[\varphi]}\,.
\end{eqnarray}
When the lattice contains only one lattice site and the averaging function $f$ is constant, the corresponding
$\Sw$ is the constraint-effective potential \cite{O'Raifeartaigh:1986hi}. 

At linear order a continuum symmetry transformation, \Eq{eq:classvar}, on the r.~h.~s.\ of \Eq{eq:SWilson} leads to the relation
\begin{eqnarray}\label{eq:transformedaction}
e^{\Sw[\phi]}\Big\langle \int \!dx\, \dec\varphi^i(x)\, \frac{\delta\R[\varphi,\phi]}{\delta\varphi^i(x) }\Big\rangle_\R -\langle \Tr \Mc\rangle_\R=0\, ,
\end{eqnarray}
where the expectation values at fixed blocked fields is defined as
\begin{equation}
 \langle \mathcal{O} \rangle_\R:=\Normw \int\, d \varphi\, e^{-\R[\varphi,\phi]-\Sc[\varphi]}\, \mathcal{O}[\varphi;\phi].
\end{equation}
These expectation values include a functional dependence on the
blocked fields $\phi$.

The continuum expectation values in relation \eq{eq:transformedaction} can be transferred into functional derivatives of the lattice fields only, provided 
the following lattice counterpart $\M$ of the continuum symmetry operator $\Mc$ can be defined, 
\begin{eqnarray}\label{eq:constraint0}
\M_{nm}^{ik} \Phi[\varphi,f]^k_m=\Phi[\Mc\varphi,f]^i_n\, .
\end{eqnarray} 
In \cite{Bergner:2008ws} this has been named `additional constraint' and its importance has been discussed for chiral symmetry and SUSY. 

Then, as derived in \cite{Bergner:2008ws}, the relation \eq{eq:transformedaction} translates into
\begin{eqnarray}\label{eq:SymWilson}
\M_{nm}^{ij}\phi_m^j \0{\delta \Sw
}{\delta \phi_n^i} & = &
(\M \alpha^{-1})_{nm}^{ij} \left( \0{ \delta
\Sw}{\delta \phi_m^j}\0{\delta \Sw }{\delta \phi_n^i}- \0{ \delta^2
\Sw}{\delta \phi_m^j\delta \phi_n^i}\right)\nonumber\\ 
&&+\Str
\M-\langle \Str\Mc \rangle_\R\,.
\end{eqnarray}
It will be interpreted as an invariance of the lattice action $\Sw$
and named `symmetry relation' or `WT identity' in due course.  

All actions
$\Sw$ defined via the blocking \eq{eq:SWilson} will automatically
fulfil the relation \eq{eq:SymWilson}. The opposite does not apply since the symmetry relation is but one of the functional relations a blocked action satisfies.

Note that the l.~h.~s.\ of this equation multiplied by the symmetry
parameter $\ep$ is just the variation of $\Sw$ to leading order
$O(\ep)$
\begin{eqnarray}
 \delta \Sw=\delta\phi^i_n\0{\delta \Sw}{\delta \phi_n^i}
=\ep\M_{nm}^{ij}\phi_m^j \0{\delta \Sw}{\delta \phi_n^i}\,.
\end{eqnarray}
Thus the r.~h.~s.\ of the symmetry relation \eq{eq:SymWilson} modifies the naive symmetry $\delta S=0$. The first line of the r.~h.~s.\ of \eq{eq:SymWilson} is independent of the averaging function $f$, rather caused by the blocking kernel $\alpha$. 
To be precise, the nonsymmetric part of $\alpha$ generates this term, it is absent for symmetric kernels \cite{Bergner:2008ws} 
or delta-blocking $\alpha^{-1}\to 0$, cf.\ \Eq{eq:SWilsondelta}.

Specialising to the case of chiral symmetry with fermions $\psi,\bpsi$
as fields, the Dirac operator $\bpsi D\psi$ as a quadratic action in
them and symmetries $\M,\Mc\propto\gamma_5$ trivially fulfilling the
additional constraint, the first and second line of the symmetry
relation \eq{eq:SymWilson} are nothing but the GW relation and the
index theorem, respectively \cite{Bergner:2008ws}. Solutions of this
relation such as the overlap operator \cite{Neuberger:1997fp} represent chiral fermions on the lattice. Its locality has been shown for not too rough configurations \cite{Hernandez:1998et}. 

Our hope is to represent SUSY on the lattice in a similar manner through actions that obey the symmetry relation  \eq{eq:SymWilson} and approach the original (classical) action in the continuum limit.

The symmetry may be looked at in a slightly different way, namely by
defining (`deformed') \emph{field-dependent} symmetry transformations for the
fields,
\begin{equation} \label{eq:ModSym}
(\M_{{\rm def}})_{nm}^{ij}\phi_m^j =
\M_{nm}^{ij}\left(\phi_m^j- (\alpha^{-1})^{jk}_{mr}\0{ \delta
\Sw}{\delta \phi_r^k} \right)\,,
\end{equation} 
which are generalisations of the modified chiral transformations on
the lattice \cite{Luscher:1998pqa}. The WT identity \eq{eq:SymWilson}
then reduces to
\begin{align}\label{eq:MdefSymWilson} 
  (\M_{{\rm def}})_{nm}^{ij}\phi_m^j \0{\delta \Sw }{\delta \phi_n^i}&
  =(-1)^{|\phi^i||\phi^j|} \0{ \delta}{ \delta \phi_n^i}\! \left[
    (\M_{\rm def})_{nm}^{ij}\phi_m^j \right]\nonumber\\&\qquad\quad
  \qquad-\Str\Mc\,.
\end{align}

Note that for Wilson fermions -- which are ultralocal but do not obey
the naive chiral symmetry, as governed by the Nielsen-Ninomiya theorem
\cite{Nielsen,Jahn:2002kg} -- one can still write down a deformed
symmetry, which, however, is non-local \cite{Bietenholz}.

\subsection{A short note on trivial solutions}
\label{sec:trivsol}
The relation \Eq{eq:SymWilson} 
leads to a deformed lattice symmetry that contains only derivatives
with respect to the lattice fields.  The only direct reference to the
continuum is encoded in \Eq{eq:constraint0}.  In order to find a
different way to relate the lattice expression with the continuum we
turn back to an intermediate step in the derivation of
\Eq{eq:SymWilson}.

The regulator \eq{eq:BlockRegulator} leads to the following relation
 of the expectation values
\begin{equation}
\Big\langle \int\!dx\,\dec\varphi^i(x) \frac{\delta\R[\varphi,\phi]}{\delta\varphi^i(x) }\Big\rangle_\R
= \Big\langle \Phi[\dec\varphi,f]^i_m\alpha^{ij}_{nm} (\Phi[\varphi,f] -\phi)_n^i \Big\rangle_\R
\end{equation}
or, equivalently,
\begin{equation}
\Big\langle \int\!dx\,\dec\varphi^i \frac{\delta\R[\varphi,\phi]}{\delta\varphi^i(x) }\Big\rangle_\R= (-1)^{|\phi^i||\dec\varphi^i|} \frac{\delta}{\delta \phi_m^i} \Big\langle \Phi[\dec\varphi,f]^i_m \Big\rangle_\R .
\label{eq:transformedaction2}
\end{equation}

In some cases the functional dependence of the r.~h.~s.\ of \Eq{eq:transformedaction2} on the lattice field $\phi$ can be solved.
In a theory without interactions the expectation value can be obtained from a saddle point approximation.
The action has in this case the simple form
\begin{equation}
 \Sc=\frac{1}{2}\int\!dx\, \varphi \tilde{K} \varphi,
\end{equation}
and the expectation value from of the saddle point solution, $\varphi_0$, of the path integral leads to
\begin{eqnarray}
\hspace{-0.5cm}\Big\langle \Phi[\dec\varphi,f]^i_m \Big\rangle_\R &=& e^{-\Sw} \ep f \Mc \varphi_0 \nonumber \\
&=&e^{-\Sw} \ep \left[ f \Mc (f^T \alpha f+\tilde{K})^{-1} f \alpha \phi\right]  .
\end{eqnarray}
In this short hand notation the application of the averaging is represented as an application of $f$ (i.~e.\ $f\varphi:=\Phi[\varphi,f]$), cf.\ 
\cite{Bergner:2008ws} for details. Hence $\ep f\Mc \varphi_0$ stands for the average of the variation of $\varphi_0$.
With this simple solution the relation \Eq{eq:transformedaction} now becomes
\begin{eqnarray}
&&e^{S[\phi]} (-1)^{|\phi^i||\dec\varphi^i|} \frac{\delta}{\delta \phi_m^i} \Big\langle \Phi[\dec\varphi,f]^i_m \Big\rangle_\R -\ep\langle \Str \Mc \rangle_\R  \nonumber\\
&&=\ep\left[ f \Mc (f^T \alpha f+\tilde{K})^{-1} f \alpha\right]^{ij}_{mn} \phi^j_n   \frac{\delta\Sw}{\delta \phi_m^i}\label{eq:trivalModSym} \\
&&+\ep \Str \left[ f \Mc (f^T \alpha f+\tilde{K})^{-1} f \alpha \right] -\ep \langle \Str \Mc \rangle_\R=0 \nonumber
\end{eqnarray}
Since the relation is linear the matrix $f \Mc (f^T \alpha
f+\tilde{K})^{-1} f \alpha$ can be interpreted as a symmetry generator
on the lattice.  Hence \Eq{eq:trivalModSym} represents a symmetry
relation on the lattice.  The main difference to the modified symmetry
relation \eq{eq:ModSym} is that $\tilde{K}$ encodes a direct reference
to the continuum action.  Such a reference seems to be unavoidable in
a solution of the expectation values in \eq{eq:transformedaction}.  in
the generic case these expectation values can can not even be solved
and only be approximations of them are available.  In our approach the
constraint in \eq{eq:constraint0} allows to avoid the reference to the
continuum action and to express the expectation values in terms of
derivatives of lattice fields. In turn, keeping the reference to the
continuum action, the additional constraint can be evaded.

Note that in a theory without interactions the lattice action
defined in \eq{eq:SWilson} follows directly by solving the Gau\ss ian
path integral.  It is the perfect action as mentioned in
\cite{Bietenholz:1996pf}.  This action is a solution of the lattice
symmetry relations \Eq{eq:trivalModSym} and \Eq{eq:ModSym}.  Apart
from the trace part the symmetry relation \Eq{eq:trivalModSym} of the
perfect action has already been found in \cite{Hasenfratz:2006kv}.

\subsection{Supersymmetric Quantum Mechanics in the Continuum}
\label{sect:susyqm}
Supersymmetric Quantum Mechanics (SUSYQM) is one of the simplest
supersymmetric models and thus ideal for investigating SUSY on the
lattice. It is a one-dimensional theory of a real boson $\tilde{\chi}$, a
bosonic auxiliary field $\tilde{F}$, a complex fermion (Grassmannian) $\tilde{\psi}$
and its complex conjugate $\tilde{\bpsi}$, which we collect into the field
vector $\varphi^i=(\tilde{\chi},\tilde{F},\tilde{\bar{\psi}},\tilde{\psi})$. The off-shell
action
\begin{eqnarray}\label{eq:susyqmaction_cont}
  \Sc[\varphi]&=&\int \!d x\,\Lc\left(\varphi(x)\right)\\
  \Lc&=&\half (\pa_x\tilde{\chi})^2
  +\tilde{\bpsi}\partial_x\tilde{\psi}-\half \tilde{F}^2
  +\tilde{\bpsi}\, \frac{\partial W}{\partial \tilde{\chi}}\,\tilde{\psi}-\tilde{F}W\nonumber\,,
\end{eqnarray}
consists of kinetic terms (actually algebraic for the auxiliary field)
and particular potential terms defined by the superpotential
$W(\tilde{\chi})$. The latter encodes mass terms and interactions,
for instance through the choice $W(\tilde{\chi})=m\tilde{\chi}^2+g\tilde{\chi}^3$.

This action is invariant under
continuum supersymmetry transformations $\dec\tilde{\chi}=-\bar{\ep}\tilde{\psi}+\ep\tilde{\bpsi}$ etc.\
up to a surface term, which we collect into
\begin{eqnarray}\label{eq:susyqm_Ms}
  \!\!\!\Mc
  =\!\left(\!\!\begin{array}{c c c c}
      0&0&0&1\\
      0&0&0&-\partial_x\\
      -\partial_x&-1&0&0\\
      0&0&0&0
\end{array}\!\!\right)\!\!,
\,
\bar{\Mc}
=\!\left(\!\begin{array}{c c c c}
    0&0&-1&0\\
    0&0&-\partial_x&0\\
    0&0&0&0\\
\partial_x&-1&0&0
\end{array}\!\right)\!\!,
\end{eqnarray}
and obviously $\left\{\Mc,\bar{\Mc}\right\}=2\partial_x$.

Let us parametrise the blocking first through its inverse as
\begin{eqnarray}
 \label{eq:redAlm}
 a(\alm)^{ij}_{mn}=
\left(
\begin{array}{cccc}
a_2 & 0 & 0 & 0 \\
0 & a_0 & 0 & 0 \\
 0 & 0 & 0 & a_1 \\
 0 & 0 & -a_1 & 0
\end{array}
\right)_{mn}\,.
\end{eqnarray}
The index of the parameters has been chosen according to the length
dimension.  This kernel can be shown to be the most general one up to
$\alpha$'s that do not contribute to the symmetry relation, cf.\ App.\
F of \cite{Bergner:2008ws}. The original $\al$ is obviously
\begin{eqnarray}
 \0{1}{a}\,\al^{ij}_{mn}=
\left(
\begin{array}{cccc}
b_2 & 0 & 0 & 0 \\
0 & b_0 & 0 & 0 \\
 0 & 0 & 0 & -b_1 \\
 0 & 0 & b_1 & 0
\end{array}
\right)_{mn}\, ,
\end{eqnarray}
where $b_{0,1,2}=1/a_{0,1,2}$.

\section{Blocking in the continuum}
\label{sec:blocking_continuum}

Our formalism can also be used to obtain equivalent theories in the
continuum, which we explore in this section. The general structure of these solutions 
helps to identify the structure of possible solutions on the lattice and identify trivial
transformations, that can be neglected. An advantage of this approach is that we do 
not need to consider the additional constraint.

To that end we do not perform an averaging, but replace
\Eq{eq:average} by $\Phi=\varphi$ (formally $f$ is the delta
distribution). Treating $an$ as a continuous variable, all formulae
can be taken over with obvious modifications (integrals instead of
sums etc.). In particular, $\M=\Mc$ and no additional constraint occurs. Hence
\begin{align}\label{eq:SWilsoncont}
e^{-\Sw[\phi]}=\Sdet{}^{1/2}\alpha \int\!\!d\varphi\,
  &e^{-\int \! dx dy\,(\phi-\varphi)(x) \alpha(x,y) (\phi-\varphi)(y)/2}\nonumber \\
\times\ &e^{-\Sc[\varphi]}\,.
\end{align}
The transformed action $S$ fulfils a relation analogous to \eq{eq:SymWilson},
\begin{align}\label{eq:SymWilsonCont}
&\int\! dxdy\, \Mc^{ij}(x,y)\,\phi^j(y) \0{\delta \Sw}{\delta \phi^i(x)}=\\\nonumber 
&\int\! dxdy(\Mc \alpha^{-1})^{ij}(x,y)\left[ \0{ \delta
\Sw}{\delta \phi^j(y)}\0{\delta \Sw }{\delta \phi^i(x)}- \0{ \delta^2
\Sw}{\delta \phi^j(y)\delta \phi^i(x)}\right]\,,
 \end{align}
or, equivalently, is invariant under field-dependent deformed symmetry transformations 
\begin{align}\label{eq:MdefCont_before} 
&\int dy\, (\Mc_{\rm def})^{ij}(x,y)\,\phi^i(y)\\\notag
&=\int \! dy \, \Mc^{ij}(x,y)\left[\phi^j(y)-\int\! dz \, (\alpha^{-1})^{jk}(y,z)\0{\delta
\Sw}{\delta \phi^k(z)}\right]\,. 
\end{align} 
that can be written like \eq{eq:MdefSymWilson},
\begin{align}\label{eq:MdefCont}
  &\int\! dxdy\, (\Mc_{\rm def})^{ij}(x,y)\,\phi^j(y) \0{\delta \Sw}{\delta \phi^i(x)} = -\Str\Mc\\\nonumber
 &+ (-1)^{|\phi^i||\phi^j|}\int \! dxdy \, \0{ \delta}{ \delta \phi^i(x)}
\left[(\Mc_{\rm def})^{ij}(x,y)\,\phi^j(y)\right]\,.
\end{align}
Even though the transformation does, strictly speaking, not correspond to a blocking of the
degrees of freedom to lattice fields we still use the name blocking in due course to mark its
similarity to the blocking transformation. 

\subsection{Blocking for SUSYQM}
\label{sec:block_susy}

For SUSYQM the blocking \eq{eq:SWilsoncont} can be performed explicitly in the
auxiliary and fermionic field, since the Lagrangian
\eq{eq:susyqmaction_cont} is bilinear in this sector. For finite
$a_0$ and $a_1$ (equivalently finite $b_0$ and $b_1$) proportional to
$\delta(x-y)$ and vanishing $a_2$/diverging $b_2$ in the sense of
\eq{eq:SWilsondelta} we obtain 
\begin{widetext}
\begin{align}\label{eq:Sintegral}
 e^{-\Sw[\chi,F,\psi,\bpsi]}
 =\Norm\!\!\int \!
d\tilde{F}d\tilde{\psi}d\tilde{\bpsi}\,
\exp\left(\!-\!\int\! dx \left[\0{b_0}{2}(F-\tilde{F})^2
+b_1(\bpsi-\tilde{\bpsi})(\psi-\tilde{\psi})
+\Lc(\chi,\tilde{F},\tilde{\psi},\tilde{\bpsi})\right]\right)
\end{align}
and thus
\begin{align}\label{eq:newcontaction}
\Sw&=\int \! dx \, \left\{\half (\pa_x\chi)^2-\half\0{b_0}{b_0-1} F^2
-\0{b_0}{b_0-1}FW-\0{1}{2(b_0-1)}W^2
  +\bpsi\left[b_1-b_1^2\left(\pa_x+\0{\pa W}{\pa \chi}+b_1\right)^{-1}
\right]\psi\right\}\nonumber\\
 &-\log\det\left[\pa_x+\0{\pa W}{\pa \chi}+b_1\right]
\end{align}
\end{widetext}

This action has several interesting properties: first of all it
depends on parameters $b_{0,1}$ which in the limit $b_{0,1}\to \infty$
-- that is diverging $\alpha$ -- lead back to the original off-shell action. 
On the other hand, in the limit $b_{0,1}\to 0$, it is the
on-shell action (with the fermionic action written as determinant),
 the auxiliary and fermionic field are just
integrated out from the original action in \eq{eq:Sintegral}.

The action fulfils the continuum symmetry relation
\eq{eq:SymWilsonCont} with nonvanishing right hand side. 
It is invariant under the field dependent transformation derived from \eq{eq:MdefCont_before}, 
\begin{align}
  \delta\chi&=-\bar{\ep}\left(\psi+a_1 \0{ \delta \Sw}{\delta \bpsi}\right)
                    +\ep\left(\bpsi-a_1\0{ \delta \Sw}{\delta \psi}\right)\,,\\ 
   \delta F&=-\bar{\ep}\pa_x\left(\psi+a_1 \0{ \delta \Sw}{\delta \bpsi}\right)
                 -\ep\pa_x\left(\bpsi-a_1 \0{ \delta \Sw}{\delta \psi}\right)\,,\\
  \delta\psi&=-\ep\pa_x\chi
               -\ep \left(F-a_0\0{ \delta \Sw}{\delta F}\right)\,,\\
  \delta\bpsi&=\bar{\ep}\pa_x\chi
               -\bar{\ep} \left(F-a_0\0{ \delta \Sw}{\delta F}\right) \,.
\label{eq:symtrfdef}
\end{align}
Let us get some more intuition about this action by comparison with
the nontrivial solution in the zero mode sector worked out in
\cite{Bergner:2008ws}. When reduced to constant fields, the action
\eq{eq:newcontaction} on a space of volume $V$ becomes
\begin{align}\label{eq:actions_blocking}
\frac{S}{V} & =  -\frac{1}{b_0-1}\left[\frac{b_0}2 F^2 + b_0 F W + \frac12 W^2\right]\\
 & + \bpsi\left(b_1 - \frac{b_1^2}{\0{\pa W}{\pa \chi} + b_1}\right)\psi
 -  \log\left(\0{\pa W}{\pa \chi} + b_1\right)\nonumber\,.
\end{align}
To specialize to a term $\lambda\bpsi\chi\psi$ as in eq.~(118) of \cite{Bergner:2008ws}, we choose
\begin{equation}
 \0{\pa W}{\pa \chi}+b_1 = \frac{b_1^2}{b_1-\lambda\chi}\,.
\end{equation}
A particular solution is
\begin{equation}\label{eq:partSol}
 W = -\frac{b_1^2}\lambda\log\left(1-\frac{\lambda\chi}{b_1}\right)-b_1\chi\,.
\end{equation}
Plugged into the action this yields
\begin{align}
  \nonumber S &= -\frac{1}{b_0-1}\left[\frac{b_0}2 F^2 - b_0 F 
    \left(\frac{b_1^2}\lambda\log\left(1-\frac{\lambda\chi}{b_1}\right)+b_1
      \chi\right) \right.\nonumber\\ 
  & \left.+ \frac12 \left(\frac{b_1^2}\lambda\log\left(1-\frac{\lambda\chi}{b_1}\right)+b_1\chi\right)^2\right]\nonumber\\
   &- \log\left(\frac{b_1^2}{b_1-\lambda\chi}\right) + \lambda\bar\psi\chi\psi\,.
\end{align}
which is very similar to the bosonic part of the interacting solution in the zero mode sector given in Eq.\ (116) of \cite{Bergner:2008ws},
\begin{align}\label{eq:old_zeromode_soln}
   \nonumber &h(\chi, F) = \frac12 F^2 - \frac{1+a_0}{a_1}\chi F +\frac{a_0(1+a_0)}{2a_1^2}\chi^2\\\nonumber
&-\left(1+\frac{1+a_0}{a_1^2\lambda}F-\frac{a_0(1+a_0)}{a_1^3\lambda}\chi\right)\log(1-a_1\lambda\chi)\\
&\ \ \ \ \ \ \ \ \ \ \ \ \ \ \ \ +\frac{a_0(1+a_0)}{2a_1^4\lambda^2}(\log(1-a_1\lambda\chi))^2\,,
\end{align}
where we have set the number of lattice points and the lattice spacing to one ($N=a=1$).  

Up to a constant, this bosonic part can be obtained from blocking by
the choice
\begin{align}\label{eq:partSol2}
  W = -\frac{b_1^2}\lambda\log\left(1-\frac{\lambda\chi}{b_1}\right)-
  b_1\chi -\frac{1}{2 (1+b_0)} \tilde F\,,
\end{align}
which is necessarily $\tilde F$-dependent. Originally, $W$ was assumed
a function of the bosonic field only. In particular, the blocking
\eq{eq:Sintegral} is not guaranteed to give the action
\eq{eq:newcontaction} anymore. However, $\tilde F$ is still Gaussian
and integrating out results in the desired action.  Note that an
$\tilde F$-dependent $W$ only yields a supersymmetric action for
constant fields, because the variation of $\tilde F$ vanishes in this
case.  This marks the difference between \eq{eq:partSol2} and the
solutions of the blocked actions derived from
\eq{eq:susyqmaction_cont} where the $\tilde F$-dependent term does not
appear.

The action \eq{eq:newcontaction} derived via blocking in the auxiliary
and fermionic sector is nontrivial: It contains interactions of
different kind than those in the original action
\eq{eq:susyqmaction_cont}, it depends on additional parameters
$b_{0,1}$ and obeys nontrivial symmetries. Yet it is fully equivalent
to the original SUSYQM system (since it is connected to the latter by
a Gaussian blocking). As such it may serve as an example, in which the
same physical content including the symmetry is represented by an
unconventional action and a nonlinear symmetry relation.

The solution for constant fields can also help to understand the
general difference between actions fulfilling the symmetry relation
and actions derived from blocking. The action
\eq{eq:old_zeromode_soln} is a solution of the symmetry relation for
constant fields. Actions obtained from the blocking, on the other
hand, are of the form \eq{eq:actions_blocking}. From the different
$F^2$ coefficients one immediately infers that the action
\eq{eq:old_zeromode_soln} cannot be obtained from the blocking.

For most of the numerical simulations, however, the differences
between the actions \eq{eq:susyqmaction_cont} and
\eq{eq:newcontaction} are negligible. Usually the auxiliary field is
integrated out and fermions are replaced by the determinant of the
operator in between them. When both computations are performed one
gets up to constants,
\begin{eqnarray}\label{eq:onshell_again}
\Sw_\text{on}\!=\!\int\!\! dx\Big[\half (\pa_x\chi)^2\!+\!\half W^2\Big]
-\log\det\left[\pa_x\!+\!\0{\pa W}{\pa \chi}\right]\,,
\end{eqnarray}
from both actions, \eq{eq:newcontaction} and \eq{eq:susyqmaction_cont}. 

Technically this comes about because the fields $\tilde{F}$ are
integrated out in the blocking. For the on-shell action the new
fields, say $F$, are integrated out as well,
\begin{align}
 \int\!dF\, e^{-\Sw[F]}&=\int\!dFd\tilde{F}e^{-b_0(F-\tilde{F})^2/2-\Sc[\tilde{F}]}\\
 &=\int\!dGd\tilde{F}e^{-b_0G^2/2-\Sc[\tilde{F}]}
 \propto\int\!d\tilde{F}e^{-\Sc[\tilde{F}]}\,,\nonumber
\end{align}
such that the resulting expression is proportional to the one obtained
by integrating out exactly those fields from the original
action. Consequently, $b_0$ (and likewise the fermionic $b_1$)
parametrising the family of actions \eq{eq:newcontaction} disappeared.

The nontrivial effect of the blocking reappears in the expectation
values of the transformed fields $\psi$, $\bar{\psi}$, and $F$.  Part
of the fermionic contribution of the action has been converted to a
nonlocal term of the bosonic part. The nonlinear symmetry
\eq{eq:symtrfdef} is reflected in terms of complicated Ward identities
relating the expectation values of the transformed fields.

Discretising the action \eq{eq:onshell_again} on the lattice, one is
still confronted with all the mentioned problems of lattice SUSY.  We
have, however, gained some information about the general solutions one
might expect from a blocking transformation.  A less trivial
reformulation would be to block the boson field by virtue of a finite
$b_2$.  This, however, requires to compute a non-Gaussian path
integral, which is as difficult as solving the system itself.

\subsection{Two dimensional Wess-Zumino model}
\label{sect:wzmodel}
\newcommand{\ct}{\chi} \newcommand{\ctc}{\chi^*}
\newcommand{\Ft}{\tilde{F}} \newcommand{\Ftc}{\tilde{F}^*}
\newcommand{\Pt}{\Psi} \newcommand{\Ptb}{\bar{\Psi}} Blockings of the
auxiliary field can always be applied in a straightfoward way.
Consider for example the Wess-Zumino model in two dimensions with
complex bosonic fields $\ct$, $\Ft$ and complex two component spinors
$\Pt$, see e.g.\ \cite{Catterall:2003ae,Bergner:2007pu}. The action
can be written in the following way
\begin{align}
\begin{split}
\Sc 
=&
\int d^2 x 
\bigg[- \frac{1}{2} 
\ctc\partial^2\ct-\frac{1}{2}|\Ft|^2 +\frac12 \Ft W' + \frac12 \Ftc (W')^* \\
&\hspace{0.0truecm} 
+ \Ptb \Big( 
\cancel
{\partial}+
P_+ W'' + P_- (W'')^*  \Big)\Pt \bigg] 
\, ,
\end{split}
\end{align}
where $P_{\pm}=\frac12 (1\pm \gamma_\ast)$ and $W(\ct)$ is some polynomial of the field $\ct$.
The blocking can now be easily applied in the auxiliary field sector:
\begin{align}
 \exp\left(\frac12 \int\! dx (F^*-\Ft^*)b_0 (F-\Ft)\right)
\end{align}
This leads to the following action
\begin{align}
\begin{split}
\Sc 
=&
\int d^2 x
\bigg[- \frac{1}{2} 
\ctc\partial^2\ct-\frac{b_0}{2(1+b_0)}|\Ft|^2
+\frac{1}{2(1+b_0)} |W'|^2\\
&\hspace{0.0truecm}
+\frac{b_0}{2(1+b_0)} (\Ft W' +\Ftc (W')^*)\\
&\hspace{0.0truecm}
+ \Ptb \Big( 
\cancel{\partial}+
P_+ W'' + P_- (W'')^*  \Big)\Pt \bigg]\, .
\end{split}
\end{align}
In the limit of $b_0 \rightarrow 0$ this is the on-shell action; in
the limit of $b_0 \rightarrow \infty$ the off-shell action. The
symmetry transformations can be deduced straightforwardly and contain
the nonlinear terms like the on-shell supersymmetry
transformations. This shows how our setup can be generalized to more
than one dimension. Again the difference of the transformed (blocked)
and the original action are only relevant for expectation values of
the auxiliary field.

\section{Translational invariance blocked}\label{sec:translation}

Supersymmetry is intimately connected to the Poin\-ca\-r\'e algebra
and thus to infinitesimal translations. Typically two SUSY
transformations anticommute to a partial derivative. In this section
we consider the anticommutator of two general linear symmetry
generators and show that a corresponding symmetry relation also holds
for it. In that way one is lead from SUSY to the symmetry relation for
translations. It is easier to analyse this relation, 
but we will discuss that it
can be fulfilled only under non-standard circumstances.
  
Let $\ep_{I}\Mc_{I}$ and $\ep_{II}\Mc_{II}$ be two infinitesimal
continuum symmetries in the sense of \eq{eq:classvar}, which both
fulfill the additional constraint \eq{eq:constraint0}. It follows
straightforwardly that every polynomial of these symmetries fulfills
this constraint, too. For SUSY with its Grassmannian $\ep$'s this
means that the anticommutator of $\Mc_{I}$ and $\Mc_{II}$ fulfills the
additional constraint.

Consider the effect of a combination of the two symmetries. The
simplest way to derive the corresponding symmetry relation is to use
the fact that in the original theory infinitesimal symmetry generators
form an algebra of the associated symmetry group.  It is a necessary,
but not sufficient, condition for the symmetry relation that the
anticommutator of two symmetry transformation also fulfills a
corresponding symmetry relation.  Therefore, the commutator of these
generators is again a generator and as such it is subject to a
symmetry relation. For supersymmetry we write
\begin{eqnarray}
 \left[\ep_{I}\Mc_{I},\ep_{II}\Mc_{II}\right]
 =\ep_{I}\ep_{II}\left\{\Mc_{I},\Mc_{II}\right\}
 \equiv\ep_{I}\ep_{II}\pa
\end{eqnarray}
defining the continuum operator $\pa$. In SUSYQM we have 
\begin{eqnarray}
 \Mc_{I}=\Mc\,,\:\Mc_{II}=\bar{\Mc}\,,\:\pa=2\partial_x\,.
\end{eqnarray}
For all
supersymmetric theories one can find $\Mc$'s that anticommute to
partial derivatives wrt.\ coordinates. Therefore, we are left to
analyse the consequences of the symmetry relation for partial
derivatives.  Obviously this can be done with just one field
species. 

In the following we investigate what kind of nontrivial modifications of the translational invariance on the lattice can 
be deduced from the symmetry relation. In particular we allow for a nonlocal antihermitian part of the operator  $\nabla$.
Such a nonlocal part arises in a natural way from the additional constraint. This consequence of the constraint for
derivative operators have already been studied in \cite{Bergner:2008ws}.

Let us restrict our investigation to one real bosonic field $\phi$ with values $\phi_n$ at
lattice sites $an$. Then the symmetry relation reads,
\begin{eqnarray}\label{eq:Sym_nabla}
\nabla_{nm}\phi_m \0{\delta \Sw
}{\delta \phi_n} & = &
(\nabla \alpha^{-1})_{nm} \left( \0{ \delta
\Sw}{\delta \phi_m}\0{\delta \Sw }{\delta \phi_n}- \0{ \delta^2
\Sw}{\delta \phi_m\delta \phi_n}\right)\nonumber\\ 
&&+\Tr
\nabla-\Tr\,\partial\,.
\end{eqnarray}
As the second line of this equation is field-independent, we can focus
on the first line.

Difference operators $\nabla$ are assumed to be anti-hermitian (giving
purely imaginary eigenvalues like partial derivatives do) and real, hence
\begin{eqnarray}
 \nabla^T=-\nabla\,,\qquad \nabla_{nm}=-\nabla_{mn}
\end{eqnarray}
By construction, $\alpha$ is symmetric and so is its inverse
\begin{eqnarray}
 \left(\alpha^{-1}\right)^T=\alpha^{-1}\,,\qquad
 \left(\alpha^{-1}\right)_{nm}=\left(\alpha^{-1}\right)_{mn}\,,
\end{eqnarray}
For the product on the r.h.s.\ of the symmetry relation
\eq{eq:Sym_nabla} it follows that
\begin{eqnarray}\label{eq:nabla_alpha_aux}
 \left(\nabla\alpha^{-1}\right)^T=-\alpha^{-1}\nabla\,.
\end{eqnarray}

Translational invariant lattice operators $X$ are circulant matrices,
$X_{m,n}=X_{m-n}$, that anticommute with the matrix
\begin{eqnarray}
 P=\left(\begin{array}{cccc}
          0&1\\
          &0&1\\
          &&\ddots&1\\
          1&&&0
         \end{array}
\right)\,.
\end{eqnarray}
Consequently, all powers of $X$ including its inverse commute with $P$.

If we assume that both $\nabla$ as well as $\alpha$ and hence $\alpha^{-1}$ are circulant,
\begin{eqnarray}
 \left[\nabla,P\right]&=0\,,\qquad\quad \nabla_{n,m}&=\nabla_{n-m}\\ 
 \left[\alpha,P\right]=0\,,\: \left[\alpha^{-1},P\right]&=0\,,\quad  \left(\alpha^{-1}\right)_{nm}&=\left(\alpha^{-1}\right)_{n-m}\,,
\end{eqnarray}
then $\nabla$ and $\alpha^{-1}$ commute and \eq{eq:nabla_alpha_aux} turns into the antisymmetry
 \begin{eqnarray}
 \left(\nabla\alpha^{-1}\right)^T=-\nabla\alpha^{-1},\:
 \left(\nabla\alpha^{-1}\right)_{nm}=-\left(\nabla\alpha^{-1}\right)_{mn}\!.
\end{eqnarray}
Since the first and second derivative of the lattice action $S$
appearing on the r.h.s. of \eq{eq:Sym_nabla} are symmetric in $n$ and
$m$, it follows that the first line on the r.h.s. of the symmetry
relation vanishes. The same conclusion can be drawn for fermion
fields, with the obvious assumptions that $\alpha$ and $S$ connecting
$\bpsi$ and $\psi$ are antisymmetric.

In this case one is back at the naive symmetry relation, where the
variation of the lattice action under infinitesimal translations must
vanish. This immediately leads to nonlocal actions if $\nabla$ is nonlocal.

There are several conceivable ways out keeping a nontrivial r.h.s. of the symmetry relation:
\begin{enumerate}
 \item $\alpha^{-1}$ is chosen non-circulant;
 \item $\nabla$ is non-circulant;
 \item both $\alpha^{-1}$ and $\nabla$ are non-circulant;
 \item $\nabla$ is given a hermitian part.
\end{enumerate}

All of them are not very natural. The first three mean that the
difference operator in the lattice SUSY transformations or/and the
blocking are not translational invariant.

For option 1 we can even show that the action cannot be translational
invariant anymore. For that purpose we move to momentum space, where the
symmetry relation reads
\begin{equation}
\nabla_{p}\phi_p\frac{\partial S}{\partial\phi_p} = \nabla_p\,\left(\alpha^{-1}
\right)_{pq}\,\left(\frac{\delta S}{\delta\phi_p}\frac{\delta S}{\delta\phi_{-q}} 
- \frac{\delta^2 S}{\delta\phi_p\delta\phi_{-q}}\right)\mbox{,}
\end{equation} 
(no sum in $p$) neglecting the trace parts.  If the action $S$ is
translational invariant, the l.h.s.\ only contains field products
whose momenta sum up to zero. This is not true for the r.h.s., because
$\alpha^{-1}$ connects different momenta. Thus the two sides can only
be equal for all field values if the action is not translational
invariant.

Option 4 resembles the Wilson-Dirac operator in gauge theories, where
a hermitian part is added to the naive one in order to lift doublers
with the disadvantages of explicitly breaking chiral symmetry and
mixed hermiticity with complex eigenvalues. 

The option 4 is beyond the scope of our current setup. In the
current setup the lattice derivative operator $\nabla$ is completely
fixed by the additional constraint, cf.\ \cite{Bergner:2008ws} for
details. A hermitian part of the operator can not be added in this
case. Hence it immediately follows from the above discussion that a
local lattice version of the symmetry relation is not
possible. However, the above discussion points to possible
modifications of our approach. A systematic modification might be
possible in the context of Section \ref{sec:trivsol} or as an
approximate solution of the additional constraint. Such an
approximation can be compared with the truncations of the nonlocal
solutions of the naive symmetry relation as derived in
\cite{Schierenberg:2012pb}.

\section{Conclusions}\label{sec:sum}

We briefly summarise our findings and discuss possible extensions of
the present work.  We have implemented the modified supersymmetry
relations derived in \cite{Bergner:2008ws} within low-dimensional
interacting supersymmetric theories. We have also presented an
alternative derivation of the lattice symmetry. Its disadvantage is a
direct dependence on the underlying continuum theory.  Hence it is not
a genuine lattice symmetry. On the other hand it evades the additional
constraint introduced in \cite{Bergner:2008ws}.

Staying in the continuum, i.e.\ without the averaging, we have studied
SUSY systems that are equivalent through quadratic blockings. We have
seen that nontrivial solutions emerge already in this simple case.
In general we find nonlinear transformations and the non-polynmial solutions
already for these simple transformations. The problem raised in \cite{Bergner:2008ws}
about the non-polynomial form of the solutions is hence rather a technical than a conceptual issue.
It appears in a derivation of a blocked action similar way in the continuum and on the lattice. 

In the zero mode sector we have compared such blocked actions to
solutions of the deformed symmetry relation from
\cite{Bergner:2008ws}. We emphasise that the symmetry relation
constitutes only one of infinitely many functional relations between
correlation functions, whereas the blocking in the path integral
determines all of them. This entails that there are more solutions to
the symmetry relation than actions obtained directly through
blockings, an example of this has been given in the zero mode
sector. However, solutions of the symmetry relation imply a
supersymmetry-improved lattice action which can be the starting point
for simulations.

The goal of our investigations has been to find local solutions of the
deformed symmetry lattice SUSY.  In \cite{Bergner:2008ws} we have
shown that the modified symmetry relations are satisfied for actions
with algebraically decaying kinetic operators; hence in a strict sense
locality is broken in these cases. It is again illustrative to compare
this with the situation for chiral fermions. There, the locality of
the Dirac operator is tightly linked to the locality of the generator
of the deformed symmetry. For example, one can also derive a deformed
symmetry relation for Wilson fermions but the related generator of the
deformed symmetry is not local. In turn, the generator of chiral
symmetry for Ginsparg-Wilson fermions is local.

In the case of deformed supersymmetry the missing locality reflects the problem with the Leibniz
rule and the fact that supersymmetry transformations do not form a compact group, as they include
translations. We have investigated the symmetry relation for translations and elaborated that the
deformed symmetry for them is the trivial one unless unconventional difference operators are used.
In any case the nonlocality of the relation can not be avoided.

This immediately raises the question for the proper definition of the necessary locality or decay
properties. In our opinion this is one key issue for practical implementation of
lattice supersymmetry by means of blocked symmetry relations. 
Another way out is to derive symmetry relations that depend on the specific continuum action of the
model under consideration. As stated in Section \ref{sec:trivsol} this requires the solution of
specific expectation values. For nontrivial theories only approximations of these can be derived
and it remains questionable if the resulting symmetry relation is enough to ensure the complete
continuum symmetry.

In summary we have pushed forward the blocking-inspired approach
towards lattice supersymmetry. In our opinion the classification of
smooth (and hard) breakings of supersymmetry from the properties of
the generator of the deformed supersymmetry would pave a way towards a
practical implementation of lattice supersymmetry. In this sense we
have advanced towards posing the right question but have not achieved
an answer yet.

\section*{Acknowledgements}
This work is supported by the Helmholtz Alliance HA216/EMMI and by the DFG under BR 2872/4-2.
G.~Bergner acknowledges support by the German BMBF contract number 06FY7100.

\renewcommand{\thesection}{}
\renewcommand{\thesubsection}{\Alph{subsection}}
\renewcommand{\theequation}{\Alph{subsection}.\arabic{equation}}


\end{document}